\documentstyle[12pt]{article}
\def\beq{\begin{equation}}
\def\eeq{\end{equation}}
\def\bea{\begin{eqnarray}}
\def\eea{\end{eqnarray}}
\def\bq{\begin{quote}}
\def\eq{\end{quote}}
 
\begin{document}
\pagestyle{empty}
\begin{flushright}
{ROME prep.1250/99 \\
 hep-th/9904201}
\end{flushright}
\vspace*{5mm}
\begin{center}
{\bf THE CONFINING BRANCH OF $QCD$}
\\  
\vspace*{1cm} 
{\bf M. Bochicchio} \\
\vspace*{0.5cm}
INFN Sezione di Roma \\
Dipartimento di Fisica, Universita' di Roma `La Sapienza' \\
Piazzale Aldo Moro 2 , 00185 Roma  \\ 
\vspace*{2cm}  
{\bf ABSTRACT  } \\
\end{center}
\vspace*{5mm}
\noindent
We show that, as a consequence of a physical interpretation based on the
Abelian projection and on the $QCD$ string, four-dimensional $QCD$ confines the 
electric flux if and only if the functional integral in the fiberwise-dual 
variables admits a hyper-Kahler reduction under the action of the gauge group.
\vspace*{1cm}
\begin{flushleft}
April 1999
\end{flushleft}
\phantom{ }
\vfill
\eject

\setcounter{page}{1}
\pagestyle{plain}

\section{Introduction}

It has been conjectured long ago that quark confinement in $QCD$ would arise
because the electric flux lines are squeezed into long flux 
tubes by the condensation of the magnetic charge \cite{H1,M}.
This mechanism of confinement is known as the dual Meissner effect, since
it is dual, in the sense of electric/magnetic duality, to the confinement of 
magnetic fluxes, that arises in a type-two superconductor, because of the 
condensation of the electric charge \cite{H2}. \\
Since the electric flux lines would span two-dimensional surfaces
embedded into the four-dimensional space-time, the dual Meissner effect
leads to an effective theory of $QCD$ in terms of closed strings \cite{Po2}. \\
The absolute confinement of the electric flux requires indeed that the flux 
line cannot break into an open string. \\
More analytically, the string functional integral would arise as the string
solution of the Migdal-Makeenko equation \cite{MM,J} in the large-$N$ limit of 
$QCD$ \cite{H4, W}. 
This is the string program, that has received recently a new revival,
most notably because of the implementation in the string setting of the
zig-zag symmetry \cite{Po2}. \\
A distinctive feature of the string program is that the existence of the
string is assumed as an ansatz for the solution of the Migdal-Makeenko equation.
In fact it is difficult to see how the strings would arise directly in terms
of the functional integral over the four-dimensional gauge connections. \\
A remarkable achievement in this direction was the representation of the 
partition function
of the two-dimensional gauge theory as a sum over branched coverings
of the two-dimensional space-time \cite{GT}. These coverings are interpreted
as the string world sheets, giving evidence in favour
of the string solution of pure gauge theories. \\
Yet this representation is obtained from the exact result for the 
two-dimensional partition function, without a direct link to gauge 
configurations in the functional integral. \\
In an unrelated development, the cotangent bundle of the moduli
of holomorphic bundles on a Riemann surface has played a key role in the 
Seiberg-Witten solution of the Coulomb branch of some four-dimensional 
supersymmetric gauge theories \cite{SW}. \\
Branched covers appear in these solutions as the spectral curves of
the characteristic equation associated to a holomorphic one-form that 
labels cotangent directions to the moduli of holomorphic bundles \cite{DW}. \\
The pre-potential, the unique holomorphic function
that determines the low-energy effective action of the supersymmetric 
theory, is constructed by means of the spectral curve. \\
More precisely, certain submanifolds of the cotangent bundle, that 
correspond to moduli of representations of the fundamental group of the 
underlying two-dimensional base manifold, admit an integrable fibration
by Jacobians of branched coverings of the base two-dimensional manifold,
the Hitchin fibration, that in turn is equivalent to assign the pre-potential.
\\ 
While no direct link to physical four-dimensional fields may be attributed
to these coverings in the framework of the Seiberg-Witten solution,
a link to the string program would possibly arise, if the cotangent
bundle of unitary connections in two-dimensions could be embedded into
the four-dimensional $QCD$ functional integral. 
Such an embedding was found in \cite{MB1}. \\
It was found there that the correct variables to define this embedding are
neither the four-dimensional gauge connections, $A$, nor their dual 
variables,
$A^D$, but a partial mixing of them, that correspond to a partial 
or fiberwise duality transformation \cite{MB1}. \\
The coordinates of the cotangent bundle of unitary connections, $T^*\cal{A}$,
appear naturally as the shift $A^D=A+\Psi$ is performed for two dualized 
polarizations among the four components of the four-dimensional gauge 
connection. \\
In addition, it was found in \cite{MB2}, that there is a dense embedding into
the $QCD$ functional integral, of an elliptic fibration of the 
moduli space of parabolic $K(D)$ pairs into (an elliptic fibration of) the 
quotient of the cotangent bundle by the action of the gauge group. \\
The last space admits a Hitchin fibration by the moduli of line bundles 
over branched spectral covers, thus giving a dense embedding of these 
objects into the $QCD$ functional integral. \\
While in \cite{MB2} the integrability properties of the Hitchin fibration 
were used to reduce the problem of computing the functional integral in the
large-$N$ limit to the evaluation of the saddle-point of a certain effective 
action that contains the Jacobian of the change of variables to the collective
field of the Hitchin fibration, in this paper we shall address the 
following, 
more qualitative issue, that relates to the string program. \\ 
What is the locus in the functional integral of the confining branch of 
$QCD$, that is, what is the locus in the moduli space of parabolic $K(D)$ 
pairs, whose image by the Hitchin map contains only Riemann surfaces spanned 
by closed strings ? \\
A partial answer to this question was given in \cite{MB3}.
In \cite{MB3} a physical interpretation of the occurrence of Hitchin 
bundles in the fiberwise dual functional integral was given, in the light 
of 't Hooft concept of Abelian projection \cite{H5}. \\
This interpretation identifies the branch points of the spectral covers as 
magnetic monopoles and the parabolic points as electric charges. \\
Since confinement requires magnetic condensation and 't Hooft alternative
excludes electric condensation, the confining branch is the locus, in the 
parabolic $K(D)$ pairs, whose image by the Hitchin map has no parabolic
singularity on the spectral cover \cite{MB3}, a not completely trivial 
condition. \\
It should be noticed that this idea is in complete analogy with the 
two-dimensional case \cite{GT}, in which the partition function is localized 
on branched coverings of the base compact space-time, without parabolic points. 
In fact the occurrence of parabolic points would imply the presence
in the vacuum to vacuum amplitudes of string diagrams with the topology of open strings,
a situation that it is appropriate to the Coulomb rather than the 
confinement phase. This last statement may be exemplified thinking to a 
sphere with two parabolic points as a topological cylinder, a vacuum diagram of
an open string theory. \\ 
We will find in this paper that the confinement locus is characterized 
precisely by the condition that the residues of the Higgs current, $\Psi$, on the 
parabolic divisor be nilpotent. \\
This condition turns out to be equivalent to the existence of a 
(dense in the large-$N$ limit) hyper-Kahler reduction of the cotangent bundle
of unitary connections under the action of the gauge group. \\
The confining branch of $QCD$ is, therefore, the hyper-Kahler locus of the 
Hitchin fibration of parabolic bundles, embedded in the $QCD$ path integral
as prescribed by fiberwise duality. \\
On the other side, this is precisely the locus for which spectral 
covers with the topology of closed string diagrams, but not open ones,
occur in the functional integral. The dual mechanism of superconductivity
and the string interpretation are therefore compatible, as it should be, and
as it has been for long time believed \cite{Po2}. \\
One more comment. 
It is a rather strange fact that the same or analogue 
objects, that are used to construct the Seiberg-Witten solution of 
four-dimensional $SUSY$ theories in the Coulomb branch, appear here as giving 
rise to a physical string interpretation of the $QCD$ functional integral, 
with an associated hyper-Kahler structure but no supersymmetry. \\
In fact we think that the explanation of this fact has much to do with 
duality as opposed to supersymmetry. \\
The Seiberg-Witten solution starts from supersymmetry, through the
structure theorem for the low-energy effective action, as determined by the 
pre-potential, and ends up with a non-linear geometric realization
of the Abelian electric magnetic/duality of the effective theory in the 
Coulomb branch, in terms of a Legendre transformation of the pre-potential 
\cite{SW}. \\
We start instead from the non-Abelian duality of the microscopic theory, as 
defined by the functional integral, to gain, by 
means of fiberwise duality and the embedding of parabolic
bundles, control over the large-$N$ limit \cite{MB2} and a mathematical 
realization of the dual Meissner effect \cite{MB3} at the same time.

\section{ The nilpotent condition}

In this section we show that the spectral covers that are in the image by
the Hitchin map of parabolic $K(D)$ pairs have no parabolic divisor if and only 
if the levels of the non-hermitian moment maps are nilpotent on each point of the 
parabolic divisor. \\
This in turn is a necessary and sufficient condition for the moduli 
space of parabolic $K(D)$ pairs to admit a hyper-Kahler structure.
In \cite{K,MB2} a special name was used to characterize this closed subspace:
parabolic Higgs bundles. \\
In any case the confinement criterium of this paper explains the physical 
meaning of the hyper-Kahler structure, a mathematical condition whose 
meaning was suspected to be physically relevant but not elucidated in 
\cite{MB2}. Indeed, there it was argued that the two cases of the parabolic 
$K(D)$ pairs and of the parabolic Higgs bundles present equivalent difficulties
from the point of view of solving the large-$N$ limit, in fact differing by
contributions of order of $\frac{1}{N}$. We now argue that parabolic Higgs 
bundles correspond to the confining branch of $QCD$ in the fiberwise-dual 
variables. \\ 
The functional integral for $QCD$ in \cite{MB2} is defined in terms of the
variables $(A_z, A_{\bar z}, \Psi_z, \Psi_{\bar z})$, obtained by means of a 
fiberwise duality transformation from $(A_z, A_{\bar z}, A_u, A_{\bar u})$, 
where $(z, \bar z, u, \bar u)$ are the complex coordinates on the product of two
two-dimensional tori, over which the theory is defined. \\
$(A_z, A_{\bar z}, \Psi_z, \Psi_{\bar z})$ define the coordinates of an 
elliptic fibration of $T^* {\cal A}$, the cotangent bundle of unitary 
connections on the $(z, \bar z)$ torus with the $(u, \bar u)$ torus as a base.
\\
The set of 
pairs $(A, \Psi)$ that are solutions of the following differential equations
(elliptically fibered over the $(u, \bar u)$ torus) 
is embedded into the space of parabolic $K(D)$ pairs \cite{MB3,MB2}:
\bea
F_A-i \Psi \wedge \Psi &=& \frac{1}{|D|}\sum_p \mu^{0}_{p} \delta_p i dz 
\wedge d\bar{z} \nonumber \\
\bar{\partial}_A \psi &=& \frac{1}{|D|}\sum_p \mu_{p} \delta_p  dz 
\wedge d\bar{z}\nonumber \\
\partial_A \bar{\psi} &=& \frac{1}{|D|}\sum_p \bar{\mu}_{p} \delta_p 
d\bar{z} \wedge dz\;
\eea
where $\delta_p$ is the two-dimensional delta-function localized at $z_p$ 
and $(\mu^{0}_{p},\mu_{p},\bar{\mu}_{p})$ are the set of levels for the 
moment maps \cite{MB2}.
The space of parabolic $K(D)$ pairs consists of a parabolic bundle with a 
holomorphic connection $\bar{\partial}_{A}$ and a parabolic morphism ${\psi}$.
Eq.(1) defines a dense stratification of the functional integral over
$T^* {\cal A}$ because the set of levels is dense everywhere in function 
space, in the sense of the distributions, as the divisor $D$ gets larger and 
larger. \\\
According to Hitchin \cite{Hi}, there is a Hitchin fibration of parabolic 
$K(D)$ pairs, defined by U(1) bundles over the following spectral cover:
\bea
Det( \lambda1 - \Psi_z)=0
\eea
The spectral cover depends only from the eigenvalues of $\Psi_z$.
The condition that the spectral cover has no parabolic point is therefore
the condition that the eigenvalues of $\Psi_z$ have no poles.
We notice that the residues of the poles of $\Psi_z$ are determined
by the levels of the non-hermitian moment maps.
In fact $\Psi_z$ can be made meromorphic with residue at the point $p$
conjugated to
the level $\mu_p$ by means of a gauge transformation $G$ in the 
complexification of the gauge group, that gauges to zero the connection 
$\bar{A}_z$,
fiberwise:
\bea
&&\bar{\partial} \psi- \frac{1}{|D|}\sum_p G \mu_{p}G^{-1} \delta_p
dz \wedge d\bar{z}=0 
\nonumber \\
&&\partial \bar{\psi}- \frac{1}{|D|}\sum_p \bar{G}^{-1}\bar{\mu}_{p}\bar{G} 
\delta_p d\bar{z} \wedge dz=0 \;
\eea
From this equation it follows that the residues of the eigenvalues of $\Psi_z$
are proportional to the eigenvalues of $\mu_{p}$.
If the eigenvalues of $\psi$ have no poles on the covering, $\mu_p$ must have
zero eigenvalues and therefore must be nilpotent and vice versa, that is the
conclusion looked forward. \\
There is however an apparent puzzle. Though the eigenvalues of $\psi$ cannot
have poles on the covering if the levels of the non-hermitian moment maps are 
nilpotent, the traces of 
powers of $\Psi_z$, that are expressed through symmetric polynomials in the
eigenvalues, certainly are meromorphic functions on the torus.
How can this happen if the eigenvalues of $\psi$ have no poles on the covering?
The answer is the following, as we have found by a direct check in the
$SU(2)$ case.
If $\mu_p$ is nilpotent, the eigenvalues of $\Psi_z$ have singularities that are
not parabolic but that look in the coordinates of the $z$ torus branched 
singularities, for example $z^{-\frac{1}{2}}$. However we should remind the 
reader that the eigenvalues of $\psi$ are really differentials on the covering.
Therefore  $z^{-\frac{1}{2}}$ should be really interpreted as 
$z^{-\frac{1}{2}}dz$, that is
$d(z^{\frac{1}{2}})$, that is, in fact, smooth on a simply branched covering.
There is no singularity on the covering. \\
Yet, symmetric powers of the eigenvalues of $\Psi_z$
may have meromorphic singularities on the torus.\\
It remains to show that if the residue of $\psi$ is nilpotent the quotient is 
hyper-Kahler. This is a known result \cite{K}.
This concludes our proof. \\
In fact a slightly stronger statement holds.
If the residues of the Higgs field are nilpotent, Eq.(1) can be interpreted 
as the vanishing condition for the moment maps of the action of the compact 
$SU(N)$ gauge group on the pair $(A, \Psi)$ and on the cotangent space of 
flags \cite{L}. The quotient under the action of the compact gauge group of
the set:
\bea
&&F_A-i \Psi \wedge \Psi - \frac{1}{|D|}\sum_p \mu^{0}_{p} \delta_p i dz 
\wedge d\bar{z}=0 \nonumber \\
&&\bar{\partial}_A \psi- \frac{1}{|D|}\sum_p n_{p} \delta_p
 dz \wedge d\bar{z}=0 \nonumber \\
&&\partial_A \bar{\psi}- \frac{1}{|D|}\sum_p \bar{n}_{p} \delta_p
d\bar{z} \wedge dz=0 \;
\eea
with fixed eigenvalues of the hermitian moment map is, by a general result 
\cite{Hi2}, the same as the quotient 
defined by the complex moment maps:
\bea
&&\bar{\partial}_A \psi- \frac{1}{|D|}\sum_p n_{p} \delta_p
 dz \wedge d\bar{z}=0 \nonumber \\
&&\partial_A \bar{\psi}- \frac{1}{|D|}\sum_p \bar{n}_{p} \delta_p
d\bar{z} \wedge dz=0\;
\eea
under the action of the complexification of the gauge group.

\section{Conclusions}

Our conclusion is that if $QCD$ confines the electric charge the functional
integral in the fiberwise dual-variables defined in \cite{MB1, MB2}
must be localized on the hyper-Kahler locus of parabolic $K(D)$ pairs,
the parabolic Higgs bundles.
This space is characterized by a nilpotent residue of the Higgs current.
These are precisely the parabolic $K(D)$ pairs whose image by the
Hitchin map contain spectral covers arbitrarily branched, but with no
parabolic points. \\
The physical interpretation is that there is a monopole condensate in the 
vacuum but no electric condensate and 
only closed electric strings occur into vacuum to vacuum diagrams.

\end{document}